# Structural and magnetic properties of CeZnAl$_3$ single crystals

Qian Liu[1], Bin Shen[1], Michael Smidman[1], Rui Li[1], ZhiYong Nie[1], XiaoYan Xiao[2], Ye Chen[1], Hanoh Lee[1*] and HuiQiu Yuan[1,3*]

[1] *Center for Correlated Matter and Department of Physics, Zhejiang University, Hangzhou 310058, China*
[2] *State Key Laboratory of Silicon Materials, School of Materials Science and Engineering, Zhejiang University, Hangzhou, Zhejiang 310027, China*
[3] *Collaborative Innovation Center of Advanced Microstructures, Nanjing 210093, China*

We have synthesized single crystals of CeZnAl$_3$, which is a new member of the family of the Ce-based intermetallics Ce$T$X$_3$ ($T$ = transition metal, $X$ = Si, Ge, Al), crystallizing in the non-centrosymmetric tetragonal BaNiSn$_3$-type structure. Magnetization, specific heat and resistivity measurements all show that CeZnAl$_3$ orders magnetically below around 4.4 K. Furthermore, magnetization measurements exhibit a hysteresis loop at low temperatures and fields, indicating the presence of a ferromagnetic component in the magnetic state. This points to a different nature of the magnetism in CeZnAl$_3$ compared to the other isostructural Ce$T$Al$_3$ compounds. Resistivity measurements under pressures up to 1.8 GPa show a moderate suppression of the ordering temperature with pressure, suggesting that measurements to higher pressures are required to look for quantum critical behavior.

## 1 Introduction

Heavy fermion materials have served as important examples of strongly correlated electron systems, where the competition between the Ruderman–Kittel–Kasuya–Yosida (RKKY) and Kondo interactions give rise to a wealth of phenomena. The relative strengths of these two interactions can be tuned via non-thermal parameters such as hydrostatic pressure, doping and applied magnetic fields, and the antiferromagnetic transition can sometimes be suppressed to zero temperature at a quantum critical point, leading to quantum critical behavior as well as unconventional superconductivity and other novel phases [1-4]. Quantum criticality in ferromagnetic materials has been less frequently investigated, and ferromagnetic quantum critical points are generally not found due to the presence of first order transitions, intervening antiferromagnetic phases, or spin-glass behavior [5]. However, recent measurements showing the divergence of the Grüneisen ratio of the Yb-based ferromagnet YbNi$_4$(P$_{0.92}$As$_{0.08}$)$_2$ revealed clear evidence for a ferromagnetic quantum critical point [6], which is not avoided by any of the aforementioned mechanisms. It is therefore important to identify ferromagnetic heavy fermion compounds, where clean single crystals can be synthesized, so as to tune the system to quantum criticality.

Ce$T$X$_3$ ($T$ = transition metal, $X$ = Si, Ge, Al) are a large family of materials where a variety of magnetic, heavy fermion and quantum critical behaviors have been probed [7]. Many of these compounds crystallize in the non-centrosymmetric BaNiSn$_3$-type structure (space group *I4mm*), which is displayed in Fig. 1(a). In this tetragonal structure, inversion symmetry is broken by the absence of the mirror plane perpendicular to the $c$-axis. A particular interest of this family has been the occurrence of pressure-induced heavy fermion superconductivity in Ce(Rh,Ir)Si$_3$, and Ce(Rh,Ir,Co)Ge$_3$[8-16], in the vicinity of the suppression of antiferromagnetic order. This is notable since the lack of inversion symmetry means that the Cooper pairs may not be entirely spin singlet or triplet, but can have mixed parity [7,17], and a number of unusual superconducting properties are observed such as extremely large and anisotropic upper critical fields, and enhanced spin susceptibilities in the superconducting state [14-16].

The Ce$T$Al$_3$ compounds have also received considerable attention recently, where both CeCuAl$_3$ and CeAuAl$_3$ form in the BaNiSn$_3$-type structure and order antiferromagnetically [18,19]. In both materials, the ordered moments which lie in the *ab*-plane have a reduced magnitude likely due to Kondo screening, but in CeCuAl$_3$ there is a commensurate in-plane modulation of the moments [20], while CeAuAl$_3$ has an incommensurate spiral spin structure which propagates along the *c*-axis [21]. An interesting result from inelastic neutron scattering measurements of CeCuAl$_3$ is the observation of an additional excitation in the paramagnetic state besides the two crystalline-electric field (CEF) excitations, which was attributed to a vibron state arising due to coupling between the CEF and phonons [22]. Meanwhile CeAgAl$_3$ was reported to order ferromagnetically at a $T_C$ of around 3 K [23], but this compound was found to have a different orthorhombic crystal structure [24]. As such it is important to look for new compounds in the Ce$TX_3$ series, particularly for ferromagnetic materials with the BaNiSn$_3$-type structure. Here we report the synthesis of single crystals of a new material with this non-centrosymmetric crystal structure, CeZnAl$_3$. We find that CeZnAl$_3$ orders magnetically below $T_M$ = 4.4 K, where magnetization measurements reveal the presence of a ferromagnetic component. Measurements of electrical resistivity under pressures up to 1.8 GPa reveal that the ordering temperature decreases under pressure, but higher pressures are needed to look for the presence of quantum criticality.

## 2 Experimental methods

Single crystals of CeZnAl$_3$ were synthesized using an Al self-flux technique. Elemental Ce, Zn and Al were combined in a molar ratio of Ce:Zn:Al=1:2:8, and sealed in evacuated silica tubes. The tubes were slowly heated to 1000℃, where they were held for 3-4 hours, before being cooled slowly to 750℃ over a period of two days, and then rapidly cooled to room temperature. The Al flux was removed using a 2 mol/L NaOH solution.

Room temperature single-crystal x-ray diffraction (XRD) measurements were performed using a Smart Apex II diffractometer with Mo-K$_\alpha$ radiation. Meanwhile the powder XRD was measured using a PANalytical X'Pert MRD diffractometer with Cu-K$_{\alpha 1}$ radiation and a graphite monochromator. Resistivity and specific heat measurements down to 2 K were performed using a Quantum Design Physical Property Measurement System (PPMS). The electrical resistivity below 2 K was measured using a $^3$He cryostat, while the magnetization measurements were performed using a Magnetic Property Measurement System (MPMS; Quantum Design). The transport measurements under pressure were carried out using a piston cylinder pressure cell with Daphne oil 7373 as pressure medium. The pressure is determined by measuring the superconducting transition temperature of Pb.

## 3 Results and discussion

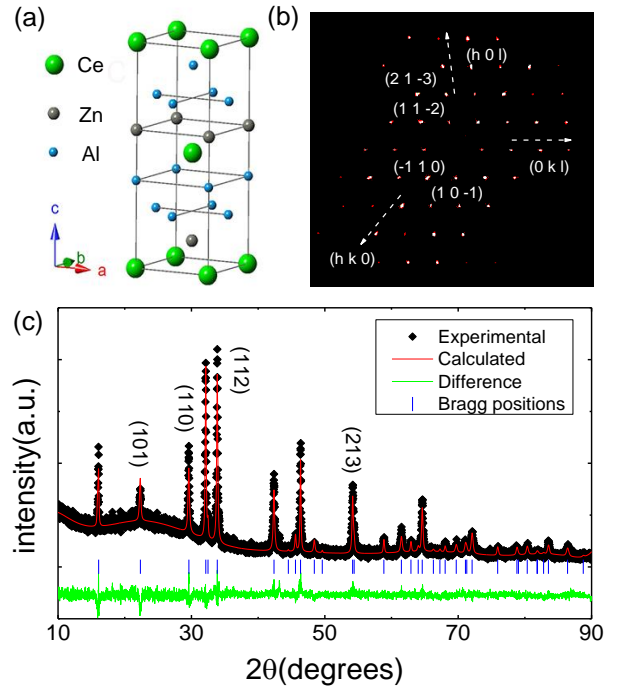

**Figure 1** (a) Crystal structure of CeZnAl$_3$. (b) Single crystal x-ray diffraction pattern of CeZnAl$_3$ measured perpendicular to the [111] direction. (c) Powder x-ray diffraction pattern of powdered CeZnAl$_3$ single crystals. The results of a Rietveld refinement, as well as the difference between the fitted curves and data are displayed by the solid lines.

The crystal structure of CeZnAl$_3$ was determined using single crystal x-ray diffraction measurements, which was found to correspond to the non-centrosymmetric BaNiSn$_3$-type structure (space group *I4mm*) displayed in Fig. 1(a). Figure 1(b) shows the single crystal XRD pattern measured perpendicular to the [111] direction, and the results of the refinement results are provided in Table 1. It can be seen that this BaNiSn$_3$-type structure can well account for the single crystal XRD results, with lattice parameters of $a$ = 4.2607(4) and $c$ = 11.004(1) Å, which are slightly larger than the isostructural CeCuAl$_3$ [24]. The powder XRD pattern for powdered CeZnAl$_3$ single crystals, as shown in Fig. 1(c), also can be well fitted using similar structural parameters to the single crystal XRD refinement.

**Table 1** Lattice constants and atomic parameters of the crystal structure of CeZnAl$_3$ from the refinement of single crystal XRD measurements.

| | Crystal system | Space group | a (Å) | c (Å) |
|---|---|---|---|---|
| Lattice | Tetragonal | I4mm | 4.2607(4) | 11.004(1) |
| Atom | Ce | Zn | Al1 | Al2 |
| X | 0.00000 | 0.00000 | 0.50000 | 0.00000 |
| Y | 0.00000 | 0.00000 | 0.00000 | 0.00000 |
| Z | 0.00000 | 0.61160 | 0.24240 | 0.38770 |
| Occ | 1.000 | 1.000 | 1.000 | 1.000 |
| $U_{11}$ | 0.0259(8) | 0.033(4) | 0.044(8) | 0.011(6) |
| $U_{22}$ | 0.0259(8) | 0.033(4) | 0.022(7) | 0.011(6) |
| $U_{33}$ | 0.0305(10) | 0.042(4) | 0.022(3) | 0.000(5) |

Residual factor R1=0.046, goodness-of-fit-on S=1.183.

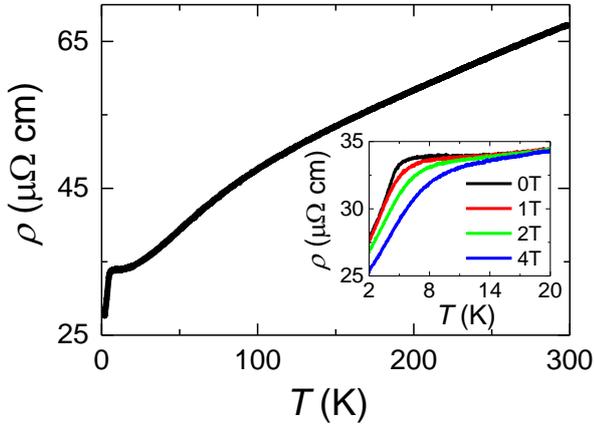

**Figure 2** Temperature dependence of the electrical resistivity $\rho(T)$ for a current perpendicular to the c-axis. The inset displays $\rho(T)$ below 20 K under various applied magnetic fields.

The main panel of Fig. 2 displays the temperature dependence of the electrical resistivity from 2 K to 300 K. The resistivity decreases upon decreasing temperature, showing metallic behavior. Below around 15 K, the resistivity is nearly constant before abruptly dropping at around 5 K, which corresponds to the onset of magnetic order. As shown in the inset, below around 20 K the overall value of the resistivity is reduced in an applied field and the anomaly at the magnetic transition becomes less pronounced and is shifted to higher temperature.

The main panel of Fig. 3(a) shows the temperature dependence of the specific heat $C(T)$ from 2 K to room temperature in zero field, and the inset shows the low temperature part. It can be seen that in zero-field, there is a sharp peak in $C(T)$ which reaches a maximum at $T_M = 4.4$ K, indicating the onset of magnetic order. The temperature dependence of the specific heat $C/T$ at various applied fields is displayed in Fig. 3(b). In fields of 1 and 2 T, the transition is suppressed to slightly lower temperatures, and the peak broadens. With further increasing magnetic field, a clear

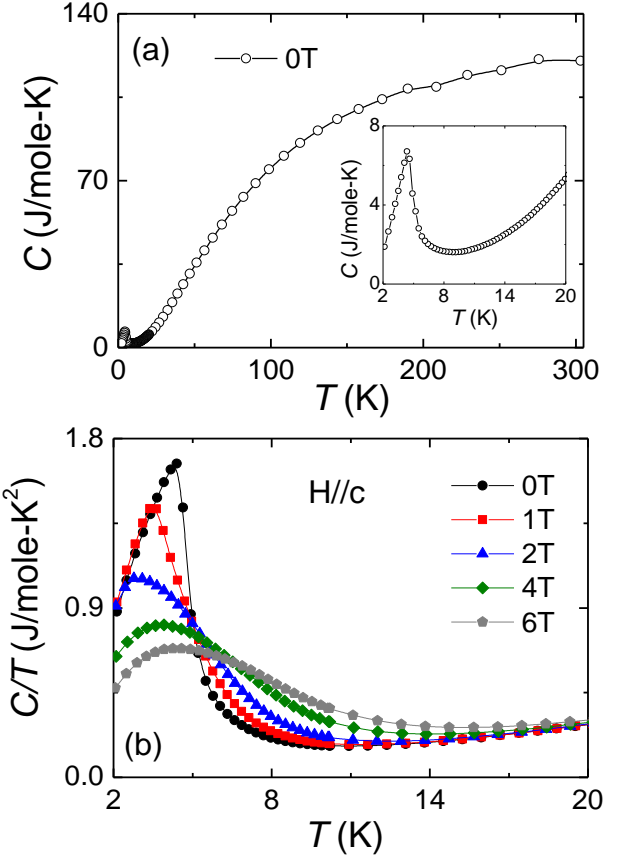

**Figure 3** (a) Temperature dependence of the specific heat $C(T)$ from 2 to 300 K. The inset displays the low temperature region. (b)Temperature dependence of the specific heat as $C(T)/T$ in various applied magnetic fields.

magnetic transition is no longer observed and the broad peak in $C/T$ shifts to higher temperature, which is commonly observed in ferromagnetic compounds.

To further probe the magnetic properties and the nature of the magnetic transition, the field dependence of the magnetization and the temperature dependence of the magnetic susceptibility were measured and are displayed in Fig. 4. At higher temperatures (Fig. 4(a)), the susceptibility shows a moderate anisotropy, the susceptibility with a field perpendicular to the c-axis ($H \perp c$) being larger than that with fields parallel ($H \| c$). Below around 8 K, this anisotropy is much more pronounced, and the susceptibility with $H \perp c$ becomes significantly enhanced. These results indicate that the easy direction corresponds to the ab-plane, similar to isostructural CeCuAl$_3$ [25]. For both field directions, the data above 100 K is well described by the Curie-Weiss expression, as shown by the plots of the inverse susceptibility in the inset of Fig. 4(a). The fitted values of the effective moment $\mu_{eff}$ and the Curie-Weiss temperature $T_{CW}$ are $\mu_{eff} = 2.74(6)$ $\mu_B$ and $T_{CW} = -7(2)$ K for $H \perp c$ and $\mu_{eff} =$

2.61(7) $\mu_B$ and $T_{CW}$ = -59(3) K for $H\|c$. The negative values of $T_{CW}$ are indicative of the presence of antiferromagnetic interactions. The deviation of the susceptibility from Curie-Weiss behavior below 100 K likely arises due to the CEF and Kondo effect. The low temperature behavior of the susceptibility with $H\perp c$ is displayed in Fig. 4(b) for both zero-field cooling (ZFC) and field-cooling (FC) curves. Below around 4.8 K it can be seen that there is a pronounced splitting of the two curves, which is a characteristic of a ferromagnetic phase. Further evidence can be seen in the field dependence of the magnetization in Fig. 4(c).

Above the ordering temperature at 8 K, there is no hysteresis in the magnetization, which increases linearly with field. Meanwhile at low temperatures in the magnetic state, a clear hysteresis loop is observed about zero-field, below around 0.1T. Such a remanent magnetization in zero-field is again a clear indication of a ferromagnetic component in the magnetic state. As shown in Fig. 4(d), the magnetization in the ordered state increases rapidly with field up to around 0.3 T, before increasing more slowly, reaching a value of around 1 $\mu_B$/Ce at 4 T.

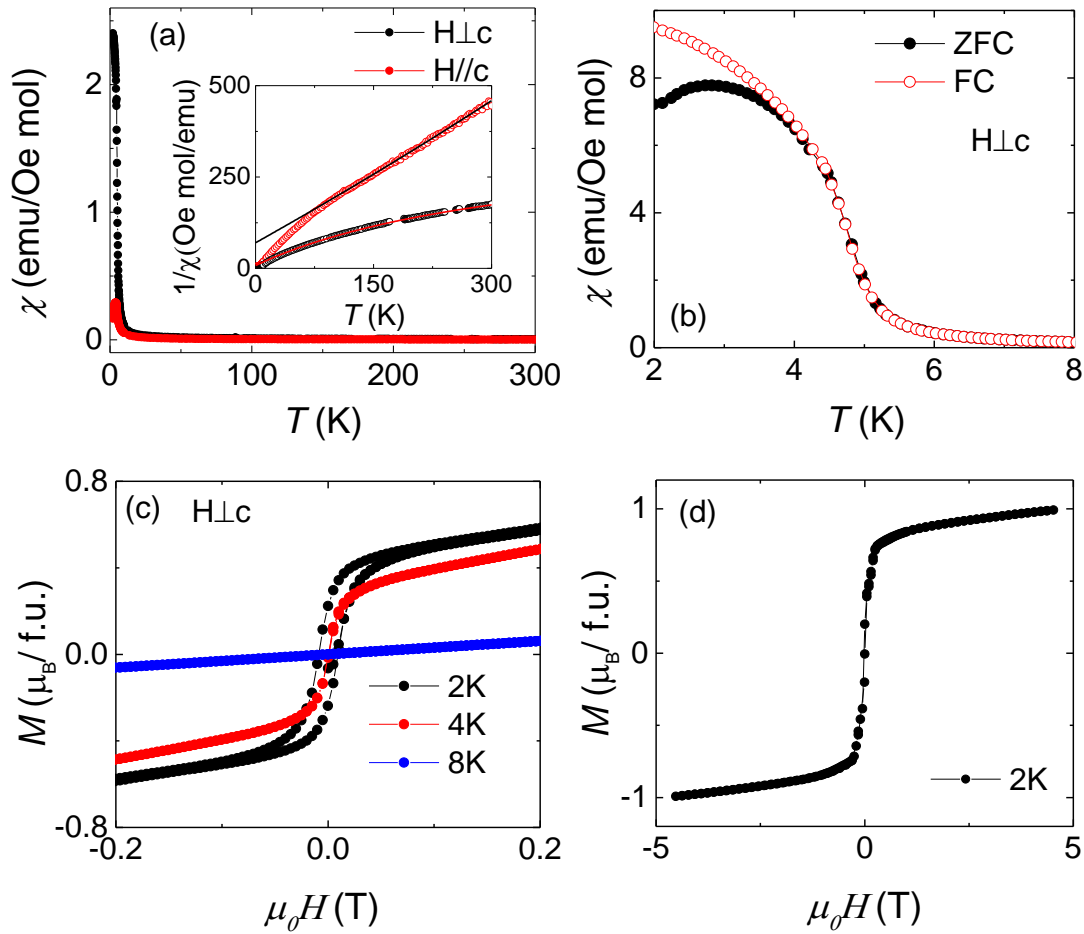

**Figure 4** (a) Temperature dependence of the magnetic susceptibility $\chi$ for magnetic fields of 0.1T applied perpendicular and parallel to the c-axis. The inset displays the inverse susceptibility against temperature, where the solid lines show the fits to the Curie-Weiss expression above 100 K. (b) Temperature dependence of the magnetic susceptibility at low temperatures for both zero-field cooling (ZFC) and field-cooling (FC) measurements. (c) Field dependence of the magnetization $M(H)$ with $H \perp c$ for low fields at $T$ = 2, 4 and 8 K, measured both sweeping the field up and down. (d) $M(H)$ over a wide field range at 2K.

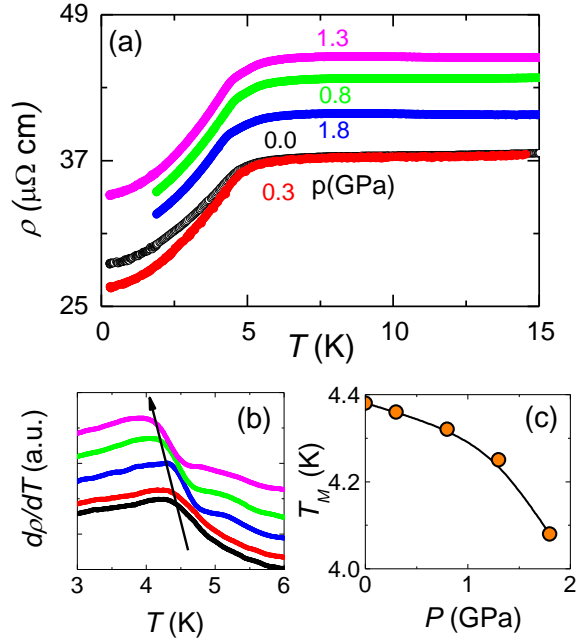

**Figure 5** (a) Temperature dependence of the electrical resistivity $\rho(T)$ under various pressures. (b) Temperature dependence of the derivative $d\rho(T)/dT$. (c) Pressure dependence of the ordering temperature $T_M$.

To examine the effects of pressure on magnetic order, and to look for possible pressure-induced quantum criticality, the temperature dependence of the resistivity was measured under pressure up to 1.8 GPa, as displayed in Fig. 5(a). It can be seen that at all pressures the presence of a magnetic transition can be found from the drop of the resistivity. By taking the magnetic transition as the position of the peak in the derivative of the resistivity (Fig. 5(b)), the magnetic ordering temperature $T_M$ is obtained as a function of pressure. This is displayed in Fig. 5(c), where $T_M$ is found to decrease slowly with increasing pressure, from 4.4K at ambient pressure to around 4.1 K at 1.8 GPa. This moderate reduction suggests that significantly larger pressures are required to look for the presence of quantum criticality and/or superconductivity.

The observation of hysteresis in magnetization loops in the magnetic state suggests the presence of ferromagnetism in CeZnAl$_3$, but whether this ordered state corresponds to a purely ferromagnetic phase, as reported for the orthorhombic CeAgAl$_3$ [23], or a state with both antiferromagnetic and ferromagnetic components needs to be determined. Indeed the latter case is found in CeCoGe$_3$ at elevated temperatures [26], as well as CeIrGe$_3$ [27]. However, these two materials show quite different magnetic properties to CeZnAl$_3$, with the ordered moments aligning strongly along the $c$-axis. On the other hand, the isostructural CeRhSi$_3$ [28], CeIrSi$_3$ [29], and CeCuAl$_3$ [25] have easy directions in the $ab$-plane. In particular, from the proposed phase diagram for Ce$TX_3$ in Ref. [21] based on the relationship of $(V_{cell})^{1/3}$ and the lattice parameter $a$, CeZnAl$_3$ would be expected to show similar magnetic properties to CeCuAl$_3$ with a purely in-plane magnetic propagation vector. Our results however, show clear differences between CeZnAl$_3$ and CeCuAl$_3$, since we reveal the presence of a ferromagnetic component at low temperatures in CeZnAl$_3$, which has not been found in CeCuAl$_3$ where there is very little splitting between ZFC and FC magnetization curves [25]. Therefore it is important to perform further measurements to characterize the magnetic structure of CeZnAl$_3$ using neutron diffraction.

## 4 Conclusion

We have synthesized single crystals of CeZnAl$_3$, a new member of the Ce$TX_3$ family, and characterized the structural and magnetic properties. CeZnAl$_3$ crystallizes in the non-centrosymmetric BaNiSn$_3$-type structure, and exhibits magnetic order at around 4.4 K, with an easy $ab$-plane. Moreover, magnetization measurements give clear evidence for a net magnetization in the ordered state, where the magnetization loops display hysteresis with a remanent magnetization in zero applied field. Additional measurements are necessary to further characterize the magnetic properties and magnetic structure of CeZnAl$_3$, particularly neutron scattering.

Measurements under pressure reveal a weak suppression of the ordering temperature up to 1.8 GPa. This is also different to the effect of pressure in CeCuAl$_3$, where the ordering temperature was found to increase with pressure up to 6 GPa, above which it could no longer be detected in resisitivity measurements [30]. As such it is also of particular interest to study at CeZnAl$_3$ at higher pressures, to examine whether there is a quantum critical point, as well as superconductivity or other novel phases.

*This work was supported by the Science Challenge Project of China (No. TZ2016004), the National Natural Science Foundation of China (No. 11474251, No. 11604291, No. U1632275), and the National Key R&D Program of China (No. 2017YFA0303100, No. 2016YFA0300202).*